\documentclass[aps,amsmath,reprint,amssymb,superscriptaddress]{revtex4-2}
\usepackage{dcolumn}
\usepackage{graphicx}
\usepackage{bm}
\usepackage{ulem}
\usepackage{color}

\usepackage{hyperref}
\begin{document}
\preprint{Version v1}
%
\title{Scaling law for the Rashba-type spin splitting in quantum well films
}
\author{Ryo~Noguchi}
\email{r-noguchi@issp.u-tokyo.ac.jp}
\affiliation{Institute for Solid State Physics (ISSP), The University of Tokyo, Kashiwa, Chiba 277-8581, Japan}

\author{Kenta~Kuroda}
\affiliation{Institute for Solid State Physics (ISSP), The University of Tokyo, Kashiwa, Chiba 277-8581, Japan}

\author{Mitsuaki~Kawamura}
\affiliation{Institute for Solid State Physics (ISSP), The University of Tokyo, Kashiwa, Chiba 277-8581, Japan}

\author{Koichiro~Yaji}
\affiliation{National Institute for Materials Science (NIMS), Tsukuba, Ibaraki 305-0003, Japan}

\author{Ayumi~Harasawa}
\affiliation{Institute for Solid State Physics (ISSP), The University of Tokyo, Kashiwa, Chiba 277-8581, Japan}

\author{Takushi~Iimori}
\affiliation{Institute for Solid State Physics (ISSP), The University of Tokyo, Kashiwa, Chiba 277-8581, Japan}

\author{Shik~Shin}
\affiliation{Institute for Solid State Physics (ISSP), The University of Tokyo, Kashiwa, Chiba 277-8581, Japan}
\affiliation{Office of University Professor, The University of Tokyo, Chiba 277-8581, Japan}

\author{Fumio~Komori}
\affiliation{Institute for Solid State Physics (ISSP), The University of Tokyo, Kashiwa, Chiba 277-8581, Japan}

\author{Taisuke~Ozaki}
\affiliation{Institute for Solid State Physics (ISSP), The University of Tokyo, Kashiwa, Chiba 277-8581, Japan}

\author{Takeshi~Kondo}
\email{kondo1215@issp.u-tokyo.ac.jp}
\affiliation{Institute for Solid State Physics (ISSP), The University of Tokyo, Kashiwa, Chiba 277-8581, Japan}
\affiliation{Trans-scale Quantum Science Institute, The University of Tokyo, Tokyo 113-0033, Japan}

\date{\today}

%
\begin{abstract}                                                   %
%
We use laser-based spin- and angle-resolved photoemission spectroscopy (laser-SARPES) with high-resolution, and experimentally determine, for the first time, the Rashba-parameters of quantum well states (QWSs) systematically changing with the film thickness and the quantum numbers, through the observation of the Ag films grown on an Au(111) substrate. The data are very well reproduced by the theoretical calculations based on the density functional theory.  Most importantly, we find a scaling law for the Rashba parameter ($\alpha_{\rm R}$) that the magnitude of $\alpha_{\rm R}$ is 
scaled by the charge density at the interface and the spin-orbit coupling ratio between the film and the substrate, and  it is expressed by a single straight line regardless of the film thickness and the quantum numbers. The new finding not only is crucial to understand the Rashba effect in QWSs but also gives a foundation of film growth engineering to fine-tune the spin splitting in 2D heterostructure systems. 
\end{abstract}
\maketitle

The manipulation of the spin split electronic state without magnetic field is one of the most important issues in the field of spintronics \cite{Soumyanarayanan2016}. In particular, the intensive researches have been given to the Rashba effect as a general property of surface or interface of solids with the strong spin-orbit coupling (SOC); this effect plays a crucial role in realizing various phenomena such as spin-charge conversion, since it can produce momentum-dependent spin-polarized electrons via the interplay between the spin-orbit coupling and the inversion asymmetry of the system \cite{bychkov1984properties,Bihlmayer2015}. 
Based on the Rashba model, the parabolic energy band of a two-dimensional electron gas with the effective mass of $m^*$ split into a pair of spin-polarized states, which are expressed as $E^{\pm} (\bm{k}) = (\hbar^2\bm{k}^2/2m^*) \pm \alpha_{\rm R} |\bm{k}|$. 
Here, $\alpha_{\rm R} $ is the Rashba parameter, and the two parabolic bands are separated by $\Delta E=2\alpha_{\rm R} |\bm{k}|$ at $k$. Through the observation of the band structure by angle-resolved photoemission spectroscopy (ARPES), the Rashba-type spin splitting has been confirmed in many systems such as the surface of heavy elements \cite{LaShell1996,Rotenberg1999,Hochstrasser2002,Hoesch2004,Koroteev2004,Hirahara2007b}, surface alloys \cite{Ast2007a,Meier2008,Moreschini2009b,Bentmann2011,Noguchi2017}, and quantum well films \cite{Dil2008,Rybkin2010}. 

\begin{figure}[t]
\begin{center}
\includegraphics[width=0.9\columnwidth]{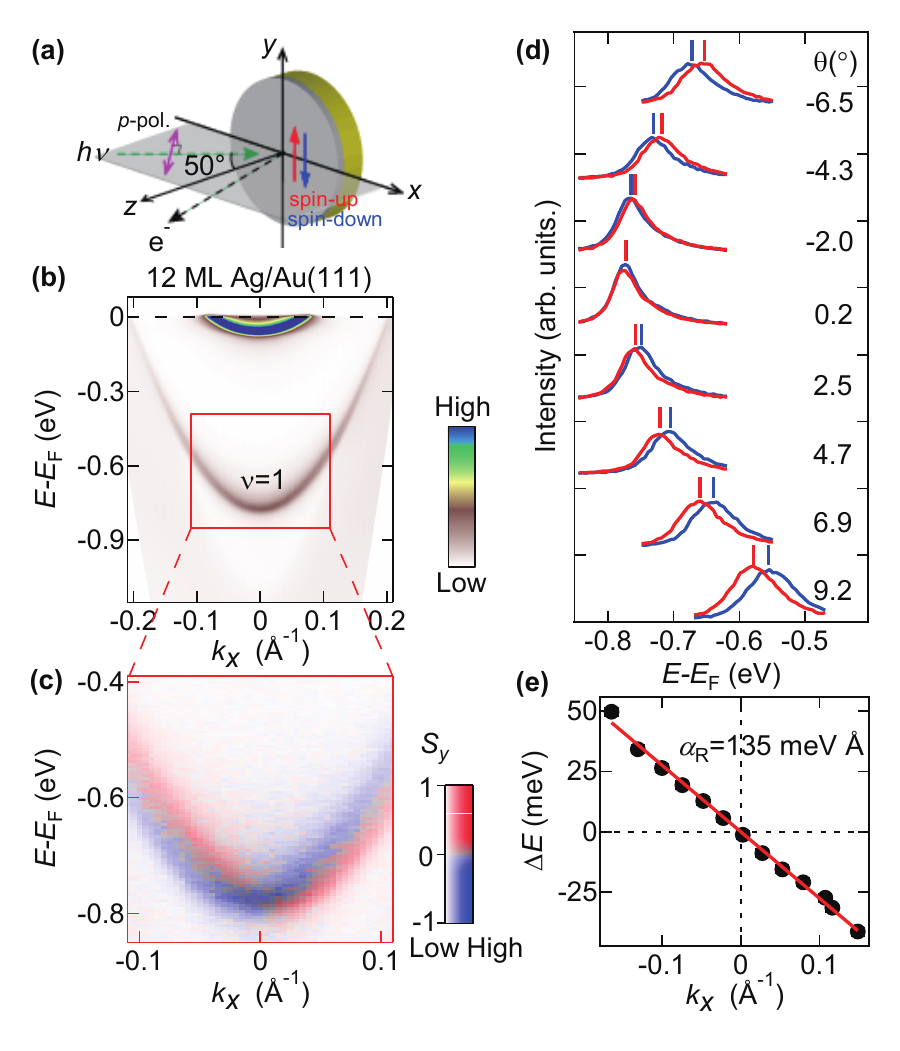}
\caption[]{(a) Experimental geometry of SARPES. (b) ARPES intensity map for 12 ML Ag/Au(111). (c) Spin-polarization and intensity map with the two-dimensional color code \cite{Tusche2015}, obtained for the red rectangle shown in (b). (d) SARPES spectra taken at different angles. The red (blue) lines show spin-up (down) spectra with peak positions indicated by red (blue) bars.  (e) Momentum-dependent energy splitting ($\Delta E$ vs $k$) and a linear fit to obtain $\alpha_{\rm R}$.}
\vspace{-10mm}
\label{fig1}
\end{center}
\end{figure}

\begin{figure*}[t]
\begin{center}
\includegraphics[width=0.8\textwidth]{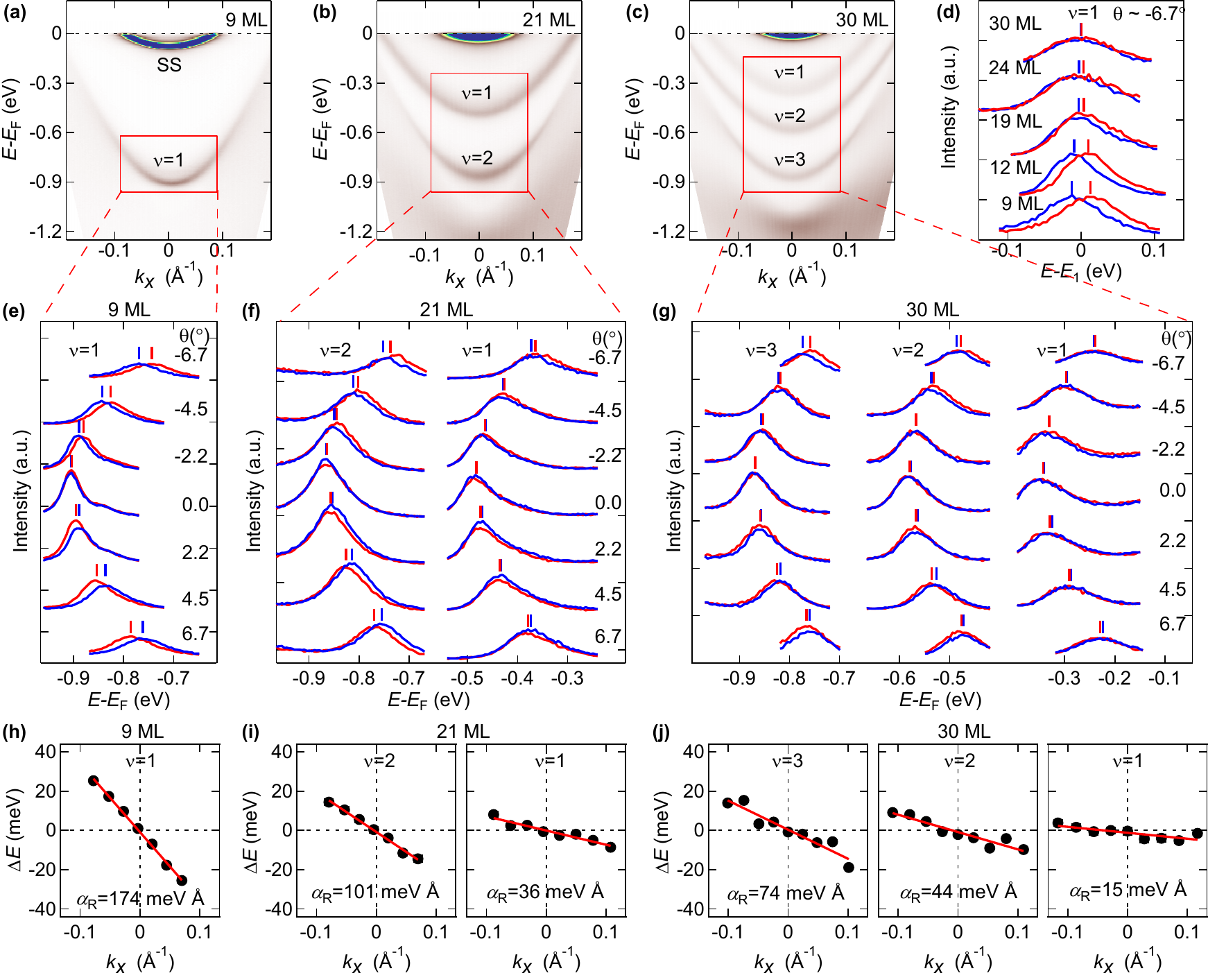}
\caption[]{(a-c) ARPES intensity maps obtained for 9, 21, 30 ML Ag/Au(111). (d) Evolution of spin-splitting of $\nu=1$ QWSs as the film thickness increases. $E_1$ is the center of the peak positions of spin-up and spin-down spectra. (e-g) Spin-resolved photoemission spectra for QWSs in 9, 21 and 30 ML Ag/Au(111). Red (Blue) bars denote the peak position of spin-up (down) spectra. (h-j) $k$-dependent energy splitting obtained from the spin-resolved spectra in (e-g). Black circles indicate the size of spin-splitting and the red lines are linear fits to obtain $\alpha_R$.}
\vspace{-5mm}
\label{fig2}
\end{center}
\end{figure*}

For 2D systems, $\alpha_{\rm R}$ is estimated as \cite{Bihlmayer2006,Nagano2009}
\begin{align}
  \alpha_{\rm R} = \int d^3 r \frac{1}{c^2}\frac{\partial V(\bf r)}{\partial z} |\psi({\bf r})|^2,
  \label{eq_rashba1}
\end{align}
where $V({\bf r})$ and $\psi({\bf r})$ are the single-body potential and the wavefunction, respectively. The $z$-axis is set to be perpendicular to the surface.
defined as vertical to the surface.  Equation (1) represents that the spin splitting is induced by the asymmetry of $V({\bm r})$ and/or $|\psi|^2$ induces,
and thus the Rashba effect can be controlled by changing these physical parameters. 
So far, the manipulation of the Rashba effect has been achieved by applying a gate voltage to quantum well states (QWSs) of semiconductor heterostructures  \cite{Engels1997,Nitta1997} or by the surface adsorption on topological insulators which generates QWSs \cite{Zhu2011a,Benia2011,King2011}; in both cases, the main origins have been attributed to the change of the potential $V(z)$. 
In contrast, the dependence of the Rashba effect on 
the charge density distribution $|\psi|^2$ has not been experimentally established to date because of the difficulty in controlling wavefunctions without applying an external field or surface band bending. 

Quantum well films offer an ideal platform to demonstrate the manipulation of the Rashba effect via tuning $|\psi|^2$
since these electrons are confined by a textbook quantum well potential with a width of the film thickness; note that the charge density distribution $|\psi|^2$ of QWSs can be controlled by changing film thicknesses. However, systematic thickness dependence of the Rashba effect has not been demonstrated until now \cite{Dil2008,Rybkin2010}, mainly due to the complex band structures of the film or the substrate. Most importantly, such a study needs to evaluate small spin splittings in the band structure, and thus it requires a high-resolution spin-resolved ARPES (SARPES), which has been developed only recently by combining the equipment with a laser photon source \cite{Yaji2016,Yaji2018}. 

To uncover the Rashba effect of QWSs, a quantum well film Ag/Au(111) is advantageous over the other QWSs systems studied so far \cite{Miller1988,McMahon1993,Chiang2000,Cercellier2004,Popovic2005,Cercellier2006,Luh2008,Forster2011}, because of the following three reasons. First, the band dispersions of QWSs formed in the projected $sp$-band gap of Au(111)  have a simple parabolic shape centered at $\bar{\Gamma}$. Secondly, the electrons of QWSs can substantially penetrate into the substrate since Ag and Au are congener and thus have similar electronic properties. Thirdly, the spin splittings of  the QWSs can get significant owing to the large SOC of Au, which is the substrate into which the QWSs penetrate. These advantages become a key for us to find the scaling law of the Rashba effect via a systematic control of the film thickness.

In this letter, we present the first demonstration on the systematic control of the Rashba effect in QWSs via tuning thicknesses of films, by investigating the band structures of Ag/Au(111) films with a high-resolution laser-SARPES. 
Furthermore, we find a scaling law governing the Rashba parameter, in which the magnitude of the spin splitting is simply expressed by the charge density at the interface, based on the agreement of the ARPES results with the density functional theory (DFT) calculations. 

The laser-ARPES and -SARPES measurements were performed at the Institute for Solid State Physics (ISSP), the University of Tokyo, using a hemispherical analyzer (ScientaOmicron, DA30L) equipped with a high-flux 6.994-eV laser~\cite{Yaji2016,Kuroda2018a}. 
The instrumental energy (angular) resolutions was set to about 5~meV (0.3$^{\circ}$) and 15~meV (0.7$^{\circ}$) for ARPES and SARPES, respectively. Ag was evaporated by a resistively heated Knudsen cell onto a clean Au(111). The film thicknesses of Ag films have been determined by the photoemission spectroscopy of quantum-well states referring to previous reports \cite{Luh2008,Forster2011}. For the theoretical analysis, we performed first-principles calculations based on DFT for slab models of Ag/Au (111). The more details of ARPES setting, sample preparation, and theoretical calculations are explained in the Supplemental Material \cite{supple}.

%

\begin{figure}[tb]
\begin{center}
\includegraphics[width=0.8\columnwidth]{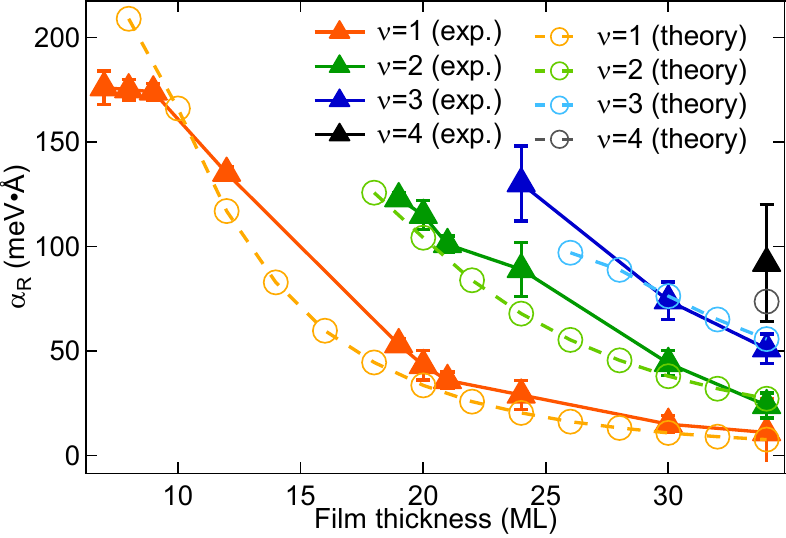}
\caption[]{Rashba parameters for $\nu=1, 2, 3, 4$ QWSs obtained by SARPES (triangles) and DFT calculations (circles) for each film thickness.
}
\vspace{-5mm}
\label{fig3}
\end{center}
\end{figure}

In Fig. \ref{fig1}(b), we plot the ARPES intensity map for the film of 12 monolayers (MLs) Ag/Au(111). The strongest intensities observed close to the Fermi level around $\bar{\Gamma}$ ($\sim$ 80 meV in the bottom) come from 
the Shockley surface, which resembles that of Ag(111) \cite{Cercellier2004}. We observed another parabolic band with the bottom at much higher binding energy ($\sim$0.8 eV) in the projected band gap of Au(111). This state is assigned to the QWS of $\nu=1$, and its envelope function is described similarly to the wavefunction of the ground state in the textbook quantum well states, which has one maximum inside the well \cite{Chiang2000}. According to theoretical studies, the QWS in a 12 ML film is not only delocalized over the film differently from the surface state, but also deeply penetrates into the Au substrate in contrast to the ideal infinite quantum well model \cite{Forster2011}. The asymmetry of the quantum well structure along the stacking direction should induce a finite spin splitting in the QWS.

To identify the spin-split bands, we performed experiments with a laser-SARPES. The SARPES image of the QWS reveals two parabolic dispersions with opposite spin polarizations [Fig. \ref{fig1}(c)], similarly to the Shockley surface states in noble metals. Some typical spectra are extracted from this image in Fig. \ref{fig1}(d), which demonstrates that the momentum dependence of spin-splitting can be determined, in high precision, from the peak positions (bars in Fig. \ref{fig1}(d)), owing to the high energy resolution of laser-SARPES; specially note that the peak position of energy distribution curve (EDC) can be determined with an accuracy that is an order of magnitude higher than the energy resolution, enabling us to estimate even a very small energy splitting of the Rashba effect. The energy splitting estimated [$\Delta E$; Fig. \ref{fig1}(d)] has a linear $k$-dependence, as expected in the Rashba model ($\Delta E=2 \alpha_{\rm R}k$). By the fitting of a linear function to the data, the Rashba parameter $\alpha_{\rm R}$ is estimated to be 135 meV${\rm \AA}$. This value is about 4 times larger than the Rashba parameter of the surface state in Ag(111) \cite{Yaji2018,supple}; this suggests that the magnitude of spin splitting in QWS is not determined simply by the surface, but to fully understand it, the asymmetry of the entire film and the substrate needs to be taken into account.

We have further studied the film thickness dependence of the spin splitting; figures \ref{fig2}(a-c) exhibit ARPES band maps for some of different film thicknesses. As the film thickness increases [from Fig. \ref{fig2}(a) to (c)], the QWSs of $\nu=1$ shift upward in energy and the QWSs with higher quantum numbers ($\nu=2,3$) emerge. Moreover, our high-quality spectra obtained with laser-SARPES reveal the peak separation of the spin-up and the spin-down spectra, which monotonically decreases with increasing film thickness [Fig. \ref{fig2}(d)]: the spin splitting at $\theta\sim-6.7^\circ$ for the QWS of $\nu$=1 is $\sim$25 meV at 9 ML, while it is $\sim$ 2 meV at 30 ML. We have confirmed that the momentum-dependent spin-texture  for all the dispersions are the same as that of 12 ML film, by tracking the peak positions of the angle-dependent SAPRES spectra exhibited in Figs. \ref{fig2}(e-g). 

\begin{figure*}[tbh]
\begin{center}
\includegraphics[width=0.9\textwidth]{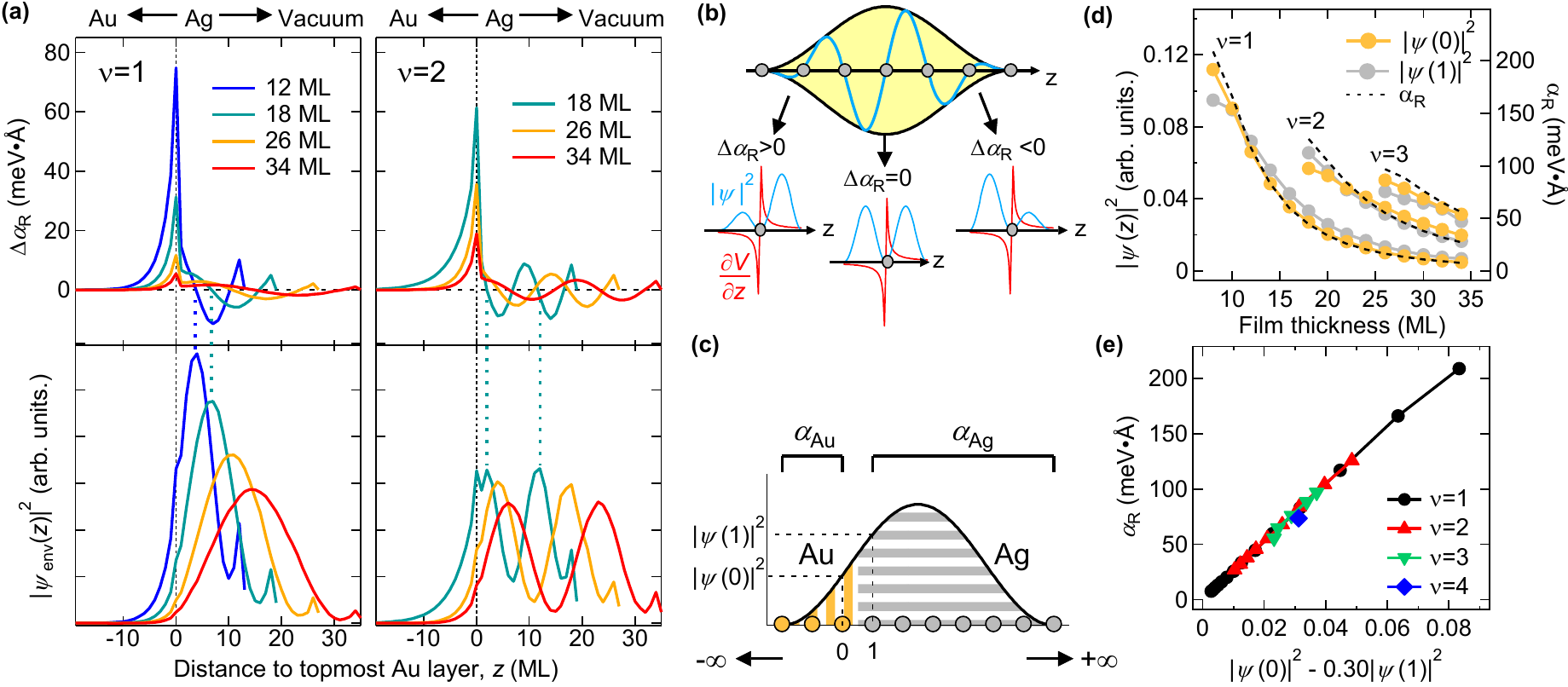}
\caption[]{(a) Contribution of each atomic layer to the spin splitting (upper panels) and the envelope functions of charge density distributions (lower panels) for $\nu=1$ QWSs (left panel) and $\nu=2$ QWSs (right panel) obtained by DFT calculations. Kink-like features in $|\psi_{env}|^2$ are due to the surface or interface effects. The blue or green dashed lines indicate the representative positions where $|\psi_{env}|^2$ has its maximum values. (b) Schematic of the relation between the charge density distribution of QWS and the layer-resolved Rashba effect. Blue curves indicate the charge density distributions and its envelope is shown by the black curve. The red curves in the lower figures denote the potential gradient $\partial V/\partial z$ which becomes larger near the nuclei. (c) Schematic of the decomposition of $\alpha_{\rm R}$ to the contributions from the Au substrate and the Ag film. (d) Local charge density distribution of QWSs at the interface Au layer ($|\psi(z_0)|^2$) and the interface Ag layer ($|\psi(z_1)|^2$) together with the magnitude of $\alpha_{\rm R}$ shown by the black dashed lines. (e) Rashba parameters of QWSs in Ag/Au(111) as a function of local charge density at the interface layer.}
\vspace{-5mm}
\label{fig4}
\end{center}
\end{figure*}

Interestingly, the spin splitting  (or $\alpha_{\rm R}$) increases for larger $\nu$ in QWSs of the same film [Figs. \ref{fig2}(h-j)]. To clarify its thicknesses dependence, we have estimated $\alpha_{\rm R}$ for films with various thicknesses and summarize these in Fig. \ref{fig3}. In this plot, a clear tendency is confirmed: when focusing on the same $\nu$ (represented by same colors), $\alpha_{\rm R}$ monotonically decreases with an increase of thickness, while it increases as $\nu$ increases at the same thickness. This is the first experimental determination of systematic thickness dependence of $\alpha_{\rm R}$ in QWSs, which thus provides us an opportunity for a reliable comparison between experiment and theory.

We found that the thickness- and $\nu$-dependence of $\alpha_{\rm R}$ obtained by DFT calculations almost perfectly reproduce our data.  The characteristic behaviors of $\alpha_{\rm R}$ could be qualitatively understood based on a finite quantum well model:  
the penetration of QWS into the substrate should be larger when the width of the well is smaller or when the quantum number $\nu$ is larger. Therefore, one can expect that QWSs in thinner Ag films or larger $\nu$s are more mixed with Au orbitals with the larger SOC, and consequently the larger spin splittings (or larger $\alpha_{\rm R}$) are realized. 

For a microscopic understanding of the Rashba effect in QWSs, we have calculated, based on DFT, the contribution of each Ag/Au layer to the Rashba parameter $\Delta \alpha_{\rm R}$ [Fig. \ref{fig4}(a)]. We find that $\Delta \alpha_{\rm R}$ of each QWS has the maximum value near the Ag/Au interface, and it oscillates within the film, even becoming negative at certain Ag layers. Moreover, the value of $\Delta\alpha_{\rm R}$ becomes 0 at the Ag layers where the envelope function has its maximum values. These results imply that the layer dependence of $\Delta\alpha_{\rm R}$ is related to the local asymmetry of the envelope function $|\psi_{env}|^2$ for the standing wave patterns of QWSs. 

The behaviors in Fig. \ref{fig4}(a) and its relation to $\alpha_{\rm R}$ would be fully understood by using mathematical formulas.
The contribution of each atom to $\alpha_{\rm R}$ can be expressed with the envelope function $\psi_{env}(z)$
as follows (see supplementary information in details for deriving the equation): 
\begin{align}
  \Delta \alpha_{{\rm R}, i}
  &\propto
  \left \langle \left| \frac{\partial V}{\partial z} \right| \right \rangle_i
  \left. \frac{\partial |\psi_{env}(z)|^2}{\partial z} \right|_{z=z_i},
\end{align}
where $z_i$ and $\langle |\partial V / \partial z| \rangle_i$ is the $z$-position and  
the gradient of the potential averaged in the vicinity of a nucleus for the $i$-th layer, respectively. As implied in Fig. \ref{fig4}(a), the sign and magnitude of $\Delta \alpha_{\rm R}$ depends on the slope of $|\psi_{env}(z)|^2$. The total Rashba splitting $\alpha_{\rm R}$ is the sum of $\Delta \alpha_{\rm R}$ across layers: $\alpha_{\rm R} = \sum_{i = -\infty}^{\infty} \Delta \alpha_{{\rm R}, i}$.
The summations could be approximated with integrals since the envelope function $|\psi_{env}(z)|^2$ varies smoothly with $z$. Furthermore, the averaged gradient $\langle |\partial V / \partial z| \rangle_i$ are two kinds for Au and Ag in the current situation [Fig.~\ref{fig4}~(c)], thus these terms can be moved outside of integral. As a result, $\alpha_{\rm R}$ is represented as follows:
\begin{align}
  \alpha_{\rm R}
  \propto&
  \left \langle \left| \frac{\partial V}{\partial z} \right| \right \rangle_{\rm Au}
  \int_{-\infty}^{0} dz
  \frac{\partial |\psi_{env}(z)|^2}{\partial z} \nonumber \\
  &+
  \left \langle \left| \frac{\partial V}{\partial z} \right| \right \rangle_{\rm Ag}
  \int_{1}^{\infty} dz
  \frac{\partial |\psi_{env}(z)|^2}{\partial z}
  \nonumber \\
  =&
  \left \langle \left| \frac{\partial V}{\partial z} \right| \right \rangle_{\rm Au}
  |\psi_{env}(0)|^2
  -
  \left \langle \left| \frac{\partial V}{\partial z} \right| \right \rangle_{\rm Ag}
  |\psi_{env}(1)|^2 \nonumber \\
  \propto&
  |\psi_{env}(0)|^2 - c_{r} |\psi_{env}(1)|^2 \label{ar}.
\end{align}
Here we have taken into account that the envelope function vanishes at $z\pm\infty$, and $c_{r}= \left \langle \left| \frac{\partial V}{\partial z} \right| \right \rangle_{\rm Ag}/\left \langle \left| \frac{\partial V}{\partial z} \right| \right \rangle_{\rm Au}$. 
This equation indicates that $\alpha_{\rm R}$ is determined only by the magnitude of charge density at the interface, and it is insensitive to the detailed structure inside the well. In addition, the ratio of the potential gradients between two atoms ($c_{r}$) plays a significant role in enhancing the Rashba effect. It is rather surprising that the Rashba effect of QWSs, a seemingly complex phenomenon, is represented in such a simple form.

To validate this mathematical analysis, the local charge densities at the interface Au/Ag layer of QWSs are obtained by calculations and compared with the Rashba parameters [Fig. \ref{fig4}(d)]. The thickness-dependence of the local charge densities at the interface [$|\psi(0)|^2$ and $|\psi(1)|^2$ in Fig. \ref{fig4}(c)] shows a similar behavior to that of $\alpha_{\rm R}$. Most importantly, we found a scaling raw for $\alpha_{\rm R}$ by plotting it against ($\psi_{env}(0)|^2 - c_{r} |\psi_{env}(1)|^2)$ with $c_{r}=0.30$, which is the ratio of the SOC coefficients of Ag $4p$ and Au $5p$ orbitals calculated in \cite{Nagano2009}. Fascinatingly, the plotted values are laid on a  single line regardless of the quantum numbers, uncovering that $\alpha_{\rm R}$ is universally scaled by the charge densities at the film interlace. 

In summary, we unveiled Rashba-type spin-splittings of QWSs in a prototype metallic quantum well system Ag/Au(111) by using high-resolution spin-resolved ARPES. The magnitude of the splitting shows a systematic variation with the film thicknesses and the quantum numbers. It was demonstrated that these data were well reproduced by the DFT calculations. Furthermore, we found the universal scaling of the Rashba parameter which is determined by the magnitude of charge densities of QWSs at the interface layers. The results provide a definitive guideline for engineering 2D heterostructures which can fine-tune the magnitude of spin-splitting toward the spintronics device application.

%
%
This work was supported by the JSPS KAKENHI (Grant No. JP18H01165, JP18K03484, JP19H02683, JP19F19030, and JP19H00651), and by MEXT Q-LEAP (Grant no. JPMXS0118068681). This work was also supported by MEXT as “Program for Promoting Researches on the Supercomputer Fugaku” (Basic Science for Emergence and Functionality in Quantum Matter Innovative Strongly Correlated Electron Science by Integration of “Fugaku” and Frontier Experiments) (Project ID: hp200132). R.N. acknowledges support by JSPS under KAKENHI Grant No. JP18J21892 and support by JSPS through the Program for Leading Graduate Schools (ALPS).
%

\begin{thebibliography}{36}%
\makeatletter
\providecommand \@ifxundefined [1]{%
 \@ifx{#1\undefined}
}%
\providecommand \@ifnum [1]{%
 \ifnum #1\expandafter \@firstoftwo
 \else \expandafter \@secondoftwo
 \fi
}%
\providecommand \@ifx [1]{%
 \ifx #1\expandafter \@firstoftwo
 \else \expandafter \@secondoftwo
 \fi
}%
\providecommand \natexlab [1]{#1}%
\providecommand \enquote  [1]{``#1''}%
\providecommand \bibnamefont  [1]{#1}%
\providecommand \bibfnamefont [1]{#1}%
\providecommand \citenamefont [1]{#1}%
\providecommand \href@noop [0]{\@secondoftwo}%
\providecommand \href [0]{\begingroup \@sanitize@url \@href}%
\providecommand \@href[1]{\@@startlink{#1}\@@href}%
\providecommand \@@href[1]{\endgroup#1\@@endlink}%
\providecommand \@sanitize@url [0]{\catcode `\\12\catcode `\$12\catcode
  `\&12\catcode `\#12\catcode `\^12\catcode `\_12\catcode `\%12\relax}%
\providecommand \@@startlink[1]{}%
\providecommand \@@endlink[0]{}%
\providecommand \url  [0]{\begingroup\@sanitize@url \@url }%
\providecommand \@url [1]{\endgroup\@href {#1}{\urlprefix }}%
\providecommand \urlprefix  [0]{URL }%
\providecommand \Eprint [0]{\href }%
\providecommand \doibase [0]{https://doi.org/}%
\providecommand \selectlanguage [0]{\@gobble}%
\providecommand \bibinfo  [0]{\@secondoftwo}%
\providecommand \bibfield  [0]{\@secondoftwo}%
\providecommand \translation [1]{[#1]}%
\providecommand \BibitemOpen [0]{}%
\providecommand \bibitemStop [0]{}%
\providecommand \bibitemNoStop [0]{.\EOS\space}%
\providecommand \EOS [0]{\spacefactor3000\relax}%
\providecommand \BibitemShut  [1]{\csname bibitem#1\endcsname}%
\let\auto@bib@innerbib\@empty
\bibitem [{\citenamefont {Soumyanarayanan}\ \emph {et~al.}(2016)\citenamefont
  {Soumyanarayanan}, \citenamefont {Reyren}, \citenamefont {Fert},\ and\
  \citenamefont {Panagopoulos}}]{Soumyanarayanan2016}%
  \BibitemOpen
  \bibfield  {author} {\bibinfo {author} {\bibfnamefont {A.}~\bibnamefont
  {Soumyanarayanan}}, \bibinfo {author} {\bibfnamefont {N.}~\bibnamefont
  {Reyren}}, \bibinfo {author} {\bibfnamefont {A.}~\bibnamefont {Fert}},\ and\
  \bibinfo {author} {\bibfnamefont {C.}~\bibnamefont {Panagopoulos}},\ }\href
  {https://doi.org/10.1038/nature19820} {\bibfield  {journal} {\bibinfo
  {journal} {Nature}\ }\textbf {\bibinfo {volume} {539}},\ \bibinfo {pages}
  {509} (\bibinfo {year} {2016})}\BibitemShut {NoStop}%
\bibitem [{\citenamefont {Bychkov}\ and\ \citenamefont
  {Rashba}(1984)}]{bychkov1984properties}%
  \BibitemOpen
  \bibfield  {author} {\bibinfo {author} {\bibfnamefont {Y.~A.}\ \bibnamefont
  {Bychkov}}\ and\ \bibinfo {author} {\bibfnamefont {E.~I.}\ \bibnamefont
  {Rashba}},\ }\href@noop {} {\bibfield  {journal} {\bibinfo  {journal} {JETP
  lett}\ }\textbf {\bibinfo {volume} {39}},\ \bibinfo {pages} {78} (\bibinfo
  {year} {1984})}\BibitemShut {NoStop}%
\bibitem [{\citenamefont {Bihlmayer}\ \emph {et~al.}(2015)\citenamefont
  {Bihlmayer}, \citenamefont {Rader},\ and\ \citenamefont
  {Winkler}}]{Bihlmayer2015}%
  \BibitemOpen
  \bibfield  {author} {\bibinfo {author} {\bibfnamefont {G.}~\bibnamefont
  {Bihlmayer}}, \bibinfo {author} {\bibfnamefont {O.}~\bibnamefont {Rader}},\
  and\ \bibinfo {author} {\bibfnamefont {R.}~\bibnamefont {Winkler}},\ }\href
  {https://doi.org/10.1088/1367-2630/17/5/050202} {\bibfield  {journal}
  {\bibinfo  {journal} {New J. Phys.}\ }\textbf {\bibinfo {volume} {17}},\
  \bibinfo {pages} {050202} (\bibinfo {year} {2015})}\BibitemShut {NoStop}%
\bibitem [{\citenamefont {LaShell}\ \emph {et~al.}(1996)\citenamefont
  {LaShell}, \citenamefont {McDougall},\ and\ \citenamefont
  {Jensen}}]{LaShell1996}%
  \BibitemOpen
  \bibfield  {author} {\bibinfo {author} {\bibfnamefont {S.}~\bibnamefont
  {LaShell}}, \bibinfo {author} {\bibfnamefont {B.~A.}\ \bibnamefont
  {McDougall}},\ and\ \bibinfo {author} {\bibfnamefont {E.}~\bibnamefont
  {Jensen}},\ }\href {https://doi.org/10.1103/PhysRevLett.77.3419} {\bibfield
  {journal} {\bibinfo  {journal} {Phys. Rev. Lett.}\ }\textbf {\bibinfo
  {volume} {77}},\ \bibinfo {pages} {3419} (\bibinfo {year}
  {1996})}\BibitemShut {NoStop}%
\bibitem [{\citenamefont {Rotenberg}\ \emph {et~al.}(1999)\citenamefont
  {Rotenberg}, \citenamefont {Chung},\ and\ \citenamefont
  {Kevan}}]{Rotenberg1999}%
  \BibitemOpen
  \bibfield  {author} {\bibinfo {author} {\bibfnamefont {E.}~\bibnamefont
  {Rotenberg}}, \bibinfo {author} {\bibfnamefont {J.~W.}\ \bibnamefont
  {Chung}},\ and\ \bibinfo {author} {\bibfnamefont {S.~D.}\ \bibnamefont
  {Kevan}},\ }\href {https://doi.org/10.1103/PhysRevLett.82.4066} {\bibfield
  {journal} {\bibinfo  {journal} {Phys. Rev. Lett.}\ }\textbf {\bibinfo
  {volume} {82}},\ \bibinfo {pages} {4066} (\bibinfo {year}
  {1999})}\BibitemShut {NoStop}%
\bibitem [{\citenamefont {Hochstrasser}\ \emph {et~al.}(2002)\citenamefont
  {Hochstrasser}, \citenamefont {Tobin}, \citenamefont {Rotenberg},\ and\
  \citenamefont {Kevan}}]{Hochstrasser2002}%
  \BibitemOpen
  \bibfield  {author} {\bibinfo {author} {\bibfnamefont {M.}~\bibnamefont
  {Hochstrasser}}, \bibinfo {author} {\bibfnamefont {J.~G.}\ \bibnamefont
  {Tobin}}, \bibinfo {author} {\bibfnamefont {E.}~\bibnamefont {Rotenberg}},\
  and\ \bibinfo {author} {\bibfnamefont {S.~D.}\ \bibnamefont {Kevan}},\ }\href
  {https://doi.org/10.1103/PhysRevLett.89.216802} {\bibfield  {journal}
  {\bibinfo  {journal} {Phys. Rev. Lett.}\ }\textbf {\bibinfo {volume} {89}},\
  \bibinfo {pages} {216802} (\bibinfo {year} {2002})}\BibitemShut {NoStop}%
\bibitem [{\citenamefont {Hoesch}\ \emph {et~al.}(2004)\citenamefont {Hoesch},
  \citenamefont {Muntwiler}, \citenamefont {Petrov}, \citenamefont
  {Hengsberger}, \citenamefont {Patthey}, \citenamefont {Shi}, \citenamefont
  {Falub}, \citenamefont {Greber},\ and\ \citenamefont
  {Osterwalder}}]{Hoesch2004}%
  \BibitemOpen
  \bibfield  {author} {\bibinfo {author} {\bibfnamefont {M.}~\bibnamefont
  {Hoesch}}, \bibinfo {author} {\bibfnamefont {M.}~\bibnamefont {Muntwiler}},
  \bibinfo {author} {\bibfnamefont {V.~N.}\ \bibnamefont {Petrov}}, \bibinfo
  {author} {\bibfnamefont {M.}~\bibnamefont {Hengsberger}}, \bibinfo {author}
  {\bibfnamefont {L.}~\bibnamefont {Patthey}}, \bibinfo {author} {\bibfnamefont
  {M.}~\bibnamefont {Shi}}, \bibinfo {author} {\bibfnamefont {M.}~\bibnamefont
  {Falub}}, \bibinfo {author} {\bibfnamefont {T.}~\bibnamefont {Greber}},\ and\
  \bibinfo {author} {\bibfnamefont {J.}~\bibnamefont {Osterwalder}},\ }\href
  {https://doi.org/10.1103/PhysRevB.69.241401} {\bibfield  {journal} {\bibinfo
  {journal} {Phys. Rev. B}\ }\textbf {\bibinfo {volume} {69}},\ \bibinfo
  {pages} {241401} (\bibinfo {year} {2004})}\BibitemShut {NoStop}%
\bibitem [{\citenamefont {Koroteev}\ \emph {et~al.}(2004)\citenamefont
  {Koroteev}, \citenamefont {Bihlmayer}, \citenamefont {Gayone}, \citenamefont
  {Chulkov}, \citenamefont {Bl{\"{u}}gel}, \citenamefont {Echenique},\ and\
  \citenamefont {Hofmann}}]{Koroteev2004}%
  \BibitemOpen
  \bibfield  {author} {\bibinfo {author} {\bibfnamefont {Y.~M.}\ \bibnamefont
  {Koroteev}}, \bibinfo {author} {\bibfnamefont {G.}~\bibnamefont {Bihlmayer}},
  \bibinfo {author} {\bibfnamefont {J.~E.}\ \bibnamefont {Gayone}}, \bibinfo
  {author} {\bibfnamefont {E.~V.}\ \bibnamefont {Chulkov}}, \bibinfo {author}
  {\bibfnamefont {S.}~\bibnamefont {Bl{\"{u}}gel}}, \bibinfo {author}
  {\bibfnamefont {P.~M.}\ \bibnamefont {Echenique}},\ and\ \bibinfo {author}
  {\bibfnamefont {P.}~\bibnamefont {Hofmann}},\ }\href
  {https://doi.org/10.1103/PhysRevLett.93.046403} {\bibfield  {journal}
  {\bibinfo  {journal} {Phys. Rev. Lett.}\ }\textbf {\bibinfo {volume} {93}},\
  \bibinfo {pages} {046403} (\bibinfo {year} {2004})}\BibitemShut {NoStop}%
\bibitem [{\citenamefont {Hirahara}\ \emph {et~al.}(2007)\citenamefont
  {Hirahara}, \citenamefont {Miyamoto}, \citenamefont {Matsuda}, \citenamefont
  {Kadono}, \citenamefont {Kimura}, \citenamefont {Nagao}, \citenamefont
  {Bihlmayer}, \citenamefont {Chulkov}, \citenamefont {Qiao}, \citenamefont
  {Shimada}, \citenamefont {Namatame}, \citenamefont {Taniguchi},\ and\
  \citenamefont {Hasegawa}}]{Hirahara2007b}%
  \BibitemOpen
  \bibfield  {author} {\bibinfo {author} {\bibfnamefont {T.}~\bibnamefont
  {Hirahara}}, \bibinfo {author} {\bibfnamefont {K.}~\bibnamefont {Miyamoto}},
  \bibinfo {author} {\bibfnamefont {I.}~\bibnamefont {Matsuda}}, \bibinfo
  {author} {\bibfnamefont {T.}~\bibnamefont {Kadono}}, \bibinfo {author}
  {\bibfnamefont {A.}~\bibnamefont {Kimura}}, \bibinfo {author} {\bibfnamefont
  {T.}~\bibnamefont {Nagao}}, \bibinfo {author} {\bibfnamefont
  {G.}~\bibnamefont {Bihlmayer}}, \bibinfo {author} {\bibfnamefont {E.~V.}\
  \bibnamefont {Chulkov}}, \bibinfo {author} {\bibfnamefont {S.}~\bibnamefont
  {Qiao}}, \bibinfo {author} {\bibfnamefont {K.}~\bibnamefont {Shimada}},
  \bibinfo {author} {\bibfnamefont {H.}~\bibnamefont {Namatame}}, \bibinfo
  {author} {\bibfnamefont {M.}~\bibnamefont {Taniguchi}},\ and\ \bibinfo
  {author} {\bibfnamefont {S.}~\bibnamefont {Hasegawa}},\ }\href
  {https://doi.org/10.1103/PhysRevB.76.153305} {\bibfield  {journal} {\bibinfo
  {journal} {Phys. Rev. B}\ }\textbf {\bibinfo {volume} {76}},\ \bibinfo
  {pages} {153305} (\bibinfo {year} {2007})}\BibitemShut {NoStop}%
\bibitem [{\citenamefont {Ast}\ \emph {et~al.}(2007)\citenamefont {Ast},
  \citenamefont {Henk}, \citenamefont {Ernst}, \citenamefont {Moreschini},
  \citenamefont {Falub}, \citenamefont {Pacil{\'{e}}}, \citenamefont {Bruno},
  \citenamefont {Kern},\ and\ \citenamefont {Grioni}}]{Ast2007a}%
  \BibitemOpen
  \bibfield  {author} {\bibinfo {author} {\bibfnamefont {C.~R.}\ \bibnamefont
  {Ast}}, \bibinfo {author} {\bibfnamefont {J.}~\bibnamefont {Henk}}, \bibinfo
  {author} {\bibfnamefont {A.}~\bibnamefont {Ernst}}, \bibinfo {author}
  {\bibfnamefont {L.}~\bibnamefont {Moreschini}}, \bibinfo {author}
  {\bibfnamefont {M.~C.}\ \bibnamefont {Falub}}, \bibinfo {author}
  {\bibfnamefont {D.}~\bibnamefont {Pacil{\'{e}}}}, \bibinfo {author}
  {\bibfnamefont {P.}~\bibnamefont {Bruno}}, \bibinfo {author} {\bibfnamefont
  {K.}~\bibnamefont {Kern}},\ and\ \bibinfo {author} {\bibfnamefont
  {M.}~\bibnamefont {Grioni}},\ }\href
  {https://doi.org/10.1103/PhysRevLett.98.186807} {\bibfield  {journal}
  {\bibinfo  {journal} {Phys. Rev. Lett.}\ }\textbf {\bibinfo {volume} {98}},\
  \bibinfo {pages} {186807} (\bibinfo {year} {2007})}\BibitemShut {NoStop}%
\bibitem [{\citenamefont {Meier}\ \emph {et~al.}(2008)\citenamefont {Meier},
  \citenamefont {Dil}, \citenamefont {Lobo-Checa}, \citenamefont {Patthey},\
  and\ \citenamefont {Osterwalder}}]{Meier2008}%
  \BibitemOpen
  \bibfield  {author} {\bibinfo {author} {\bibfnamefont {F.}~\bibnamefont
  {Meier}}, \bibinfo {author} {\bibfnamefont {H.}~\bibnamefont {Dil}}, \bibinfo
  {author} {\bibfnamefont {J.}~\bibnamefont {Lobo-Checa}}, \bibinfo {author}
  {\bibfnamefont {L.}~\bibnamefont {Patthey}},\ and\ \bibinfo {author}
  {\bibfnamefont {J.}~\bibnamefont {Osterwalder}},\ }\href
  {https://doi.org/10.1103/PhysRevB.77.165431} {\bibfield  {journal} {\bibinfo
  {journal} {Phys. Rev. B}\ }\textbf {\bibinfo {volume} {77}},\ \bibinfo
  {pages} {165431} (\bibinfo {year} {2008})}\BibitemShut {NoStop}%
\bibitem [{\citenamefont {Moreschini}\ \emph {et~al.}(2009)\citenamefont
  {Moreschini}, \citenamefont {Bendounan}, \citenamefont {Bentmann},
  \citenamefont {Assig}, \citenamefont {Kern}, \citenamefont {Reinert},
  \citenamefont {Henk}, \citenamefont {Ast},\ and\ \citenamefont
  {Grioni}}]{Moreschini2009b}%
  \BibitemOpen
  \bibfield  {author} {\bibinfo {author} {\bibfnamefont {L.}~\bibnamefont
  {Moreschini}}, \bibinfo {author} {\bibfnamefont {A.}~\bibnamefont
  {Bendounan}}, \bibinfo {author} {\bibfnamefont {H.}~\bibnamefont {Bentmann}},
  \bibinfo {author} {\bibfnamefont {M.}~\bibnamefont {Assig}}, \bibinfo
  {author} {\bibfnamefont {K.}~\bibnamefont {Kern}}, \bibinfo {author}
  {\bibfnamefont {F.}~\bibnamefont {Reinert}}, \bibinfo {author} {\bibfnamefont
  {J.}~\bibnamefont {Henk}}, \bibinfo {author} {\bibfnamefont {C.~R.}\
  \bibnamefont {Ast}},\ and\ \bibinfo {author} {\bibfnamefont {M.}~\bibnamefont
  {Grioni}},\ }\href {https://doi.org/10.1103/PhysRevB.80.035438} {\bibfield
  {journal} {\bibinfo  {journal} {Phys. Rev. B}\ }\textbf {\bibinfo {volume}
  {80}},\ \bibinfo {pages} {035438} (\bibinfo {year} {2009})}\BibitemShut
  {NoStop}%
\bibitem [{\citenamefont {Bentmann}\ \emph {et~al.}(2011)\citenamefont
  {Bentmann}, \citenamefont {Kuzumaki}, \citenamefont {Bihlmayer},
  \citenamefont {Bl{\"{u}}gel}, \citenamefont {Chulkov}, \citenamefont
  {Reinert},\ and\ \citenamefont {Sakamoto}}]{Bentmann2011}%
  \BibitemOpen
  \bibfield  {author} {\bibinfo {author} {\bibfnamefont {H.}~\bibnamefont
  {Bentmann}}, \bibinfo {author} {\bibfnamefont {T.}~\bibnamefont {Kuzumaki}},
  \bibinfo {author} {\bibfnamefont {G.}~\bibnamefont {Bihlmayer}}, \bibinfo
  {author} {\bibfnamefont {S.}~\bibnamefont {Bl{\"{u}}gel}}, \bibinfo {author}
  {\bibfnamefont {E.~V.}\ \bibnamefont {Chulkov}}, \bibinfo {author}
  {\bibfnamefont {F.}~\bibnamefont {Reinert}},\ and\ \bibinfo {author}
  {\bibfnamefont {K.}~\bibnamefont {Sakamoto}},\ }\href
  {https://doi.org/10.1103/PhysRevB.84.115426} {\bibfield  {journal} {\bibinfo
  {journal} {Phys. Rev. B}\ }\textbf {\bibinfo {volume} {84}},\ \bibinfo
  {pages} {115426} (\bibinfo {year} {2011})}\BibitemShut {NoStop}%
\bibitem [{\citenamefont {Noguchi}\ \emph {et~al.}(2017)\citenamefont
  {Noguchi}, \citenamefont {Kuroda}, \citenamefont {Yaji}, \citenamefont
  {Kobayashi}, \citenamefont {Sakano}, \citenamefont {Harasawa}, \citenamefont
  {Kondo}, \citenamefont {Komori},\ and\ \citenamefont {Shin}}]{Noguchi2017}%
  \BibitemOpen
  \bibfield  {author} {\bibinfo {author} {\bibfnamefont {R.}~\bibnamefont
  {Noguchi}}, \bibinfo {author} {\bibfnamefont {K.}~\bibnamefont {Kuroda}},
  \bibinfo {author} {\bibfnamefont {K.}~\bibnamefont {Yaji}}, \bibinfo {author}
  {\bibfnamefont {K.}~\bibnamefont {Kobayashi}}, \bibinfo {author}
  {\bibfnamefont {M.}~\bibnamefont {Sakano}}, \bibinfo {author} {\bibfnamefont
  {A.}~\bibnamefont {Harasawa}}, \bibinfo {author} {\bibfnamefont
  {T.}~\bibnamefont {Kondo}}, \bibinfo {author} {\bibfnamefont
  {F.}~\bibnamefont {Komori}},\ and\ \bibinfo {author} {\bibfnamefont
  {S.}~\bibnamefont {Shin}},\ }\href
  {https://doi.org/10.1103/PhysRevB.95.041111} {\bibfield  {journal} {\bibinfo
  {journal} {Phys. Rev. B}\ }\textbf {\bibinfo {volume} {95}},\ \bibinfo
  {pages} {041111} (\bibinfo {year} {2017})}\BibitemShut {NoStop}%
\bibitem [{\citenamefont {Dil}\ \emph {et~al.}(2008)\citenamefont {Dil},
  \citenamefont {Meier}, \citenamefont {Lobo-Checa}, \citenamefont {Patthey},
  \citenamefont {Bihlmayer},\ and\ \citenamefont {Osterwalder}}]{Dil2008}%
  \BibitemOpen
  \bibfield  {author} {\bibinfo {author} {\bibfnamefont {J.~H.}\ \bibnamefont
  {Dil}}, \bibinfo {author} {\bibfnamefont {F.}~\bibnamefont {Meier}}, \bibinfo
  {author} {\bibfnamefont {J.}~\bibnamefont {Lobo-Checa}}, \bibinfo {author}
  {\bibfnamefont {L.}~\bibnamefont {Patthey}}, \bibinfo {author} {\bibfnamefont
  {G.}~\bibnamefont {Bihlmayer}},\ and\ \bibinfo {author} {\bibfnamefont
  {J.}~\bibnamefont {Osterwalder}},\ }\href
  {https://doi.org/10.1103/PhysRevLett.101.266802} {\bibfield  {journal}
  {\bibinfo  {journal} {Phys. Rev. Lett.}\ }\textbf {\bibinfo {volume} {101}},\
  \bibinfo {pages} {266802} (\bibinfo {year} {2008})}\BibitemShut {NoStop}%
\bibitem [{\citenamefont {Rybkin}\ \emph {et~al.}(2010)\citenamefont {Rybkin},
  \citenamefont {Shikin}, \citenamefont {Adamchuk}, \citenamefont {Marchenko},
  \citenamefont {Biswas}, \citenamefont {Varykhalov},\ and\ \citenamefont
  {Rader}}]{Rybkin2010}%
  \BibitemOpen
  \bibfield  {author} {\bibinfo {author} {\bibfnamefont {A.~G.}\ \bibnamefont
  {Rybkin}}, \bibinfo {author} {\bibfnamefont {A.~M.}\ \bibnamefont {Shikin}},
  \bibinfo {author} {\bibfnamefont {V.~K.}\ \bibnamefont {Adamchuk}}, \bibinfo
  {author} {\bibfnamefont {D.}~\bibnamefont {Marchenko}}, \bibinfo {author}
  {\bibfnamefont {C.}~\bibnamefont {Biswas}}, \bibinfo {author} {\bibfnamefont
  {A.}~\bibnamefont {Varykhalov}},\ and\ \bibinfo {author} {\bibfnamefont
  {O.}~\bibnamefont {Rader}},\ }\href
  {https://doi.org/10.1103/PhysRevB.82.233403} {\bibfield  {journal} {\bibinfo
  {journal} {Phys. Rev. B}\ }\textbf {\bibinfo {volume} {82}},\ \bibinfo
  {pages} {233403} (\bibinfo {year} {2010})}\BibitemShut {NoStop}%
\bibitem [{\citenamefont {Tusche}\ \emph {et~al.}(2015)\citenamefont {Tusche},
  \citenamefont {Krasyuk},\ and\ \citenamefont {Kirschner}}]{Tusche2015}%
  \BibitemOpen
  \bibfield  {author} {\bibinfo {author} {\bibfnamefont {C.}~\bibnamefont
  {Tusche}}, \bibinfo {author} {\bibfnamefont {A.}~\bibnamefont {Krasyuk}},\
  and\ \bibinfo {author} {\bibfnamefont {J.}~\bibnamefont {Kirschner}},\ }\href
  {https://doi.org/10.1016/j.ultramic.2015.03.020} {\bibfield  {journal}
  {\bibinfo  {journal} {Ultramicroscopy}\ }\textbf {\bibinfo {volume} {159}},\
  \bibinfo {pages} {520} (\bibinfo {year} {2015})}\BibitemShut {NoStop}%
\bibitem [{\citenamefont {Bihlmayer}\ \emph {et~al.}(2006)\citenamefont
  {Bihlmayer}, \citenamefont {Koroteev}, \citenamefont {Echenique},
  \citenamefont {Chulkov},\ and\ \citenamefont {Bl{\"{u}}gel}}]{Bihlmayer2006}%
  \BibitemOpen
  \bibfield  {author} {\bibinfo {author} {\bibfnamefont {G.}~\bibnamefont
  {Bihlmayer}}, \bibinfo {author} {\bibfnamefont {Y.}~\bibnamefont {Koroteev}},
  \bibinfo {author} {\bibfnamefont {P.}~\bibnamefont {Echenique}}, \bibinfo
  {author} {\bibfnamefont {E.}~\bibnamefont {Chulkov}},\ and\ \bibinfo {author}
  {\bibfnamefont {S.}~\bibnamefont {Bl{\"{u}}gel}},\ }\href
  {https://doi.org/10.1016/j.susc.2006.01.098} {\bibfield  {journal} {\bibinfo
  {journal} {Surf. Sci.}\ }\textbf {\bibinfo {volume} {600}},\ \bibinfo {pages}
  {3888} (\bibinfo {year} {2006})}\BibitemShut {NoStop}%
\bibitem [{\citenamefont {Nagano}\ \emph {et~al.}(2009)\citenamefont {Nagano},
  \citenamefont {Kodama}, \citenamefont {Shishidou},\ and\ \citenamefont
  {Oguchi}}]{Nagano2009}%
  \BibitemOpen
  \bibfield  {author} {\bibinfo {author} {\bibfnamefont {M.}~\bibnamefont
  {Nagano}}, \bibinfo {author} {\bibfnamefont {A.}~\bibnamefont {Kodama}},
  \bibinfo {author} {\bibfnamefont {T.}~\bibnamefont {Shishidou}},\ and\
  \bibinfo {author} {\bibfnamefont {T.}~\bibnamefont {Oguchi}},\ }\href
  {https://doi.org/10.1088/0953-8984/21/6/064239} {\bibfield  {journal}
  {\bibinfo  {journal} {J. Phys. Condens. Matter}\ }\textbf {\bibinfo {volume}
  {21}},\ \bibinfo {pages} {064239} (\bibinfo {year} {2009})}\BibitemShut
  {NoStop}%
\bibitem [{\citenamefont {Engels}\ \emph {et~al.}(1997)\citenamefont {Engels},
  \citenamefont {Lange}, \citenamefont {Sch{\"{a}}pers},\ and\ \citenamefont
  {L{\"{u}}th}}]{Engels1997}%
  \BibitemOpen
  \bibfield  {author} {\bibinfo {author} {\bibfnamefont {G.}~\bibnamefont
  {Engels}}, \bibinfo {author} {\bibfnamefont {J.}~\bibnamefont {Lange}},
  \bibinfo {author} {\bibfnamefont {T.}~\bibnamefont {Sch{\"{a}}pers}},\ and\
  \bibinfo {author} {\bibfnamefont {H.}~\bibnamefont {L{\"{u}}th}},\ }\href
  {https://doi.org/10.1103/PhysRevB.55.R1958} {\bibfield  {journal} {\bibinfo
  {journal} {Phys. Rev. B}\ }\textbf {\bibinfo {volume} {55}},\ \bibinfo
  {pages} {R1958} (\bibinfo {year} {1997})}\BibitemShut {NoStop}%
\bibitem [{\citenamefont {Nitta}\ \emph {et~al.}(1997)\citenamefont {Nitta},
  \citenamefont {Akazaki}, \citenamefont {Takayanagi},\ and\ \citenamefont
  {Enoki}}]{Nitta1997}%
  \BibitemOpen
  \bibfield  {author} {\bibinfo {author} {\bibfnamefont {J.}~\bibnamefont
  {Nitta}}, \bibinfo {author} {\bibfnamefont {T.}~\bibnamefont {Akazaki}},
  \bibinfo {author} {\bibfnamefont {H.}~\bibnamefont {Takayanagi}},\ and\
  \bibinfo {author} {\bibfnamefont {T.}~\bibnamefont {Enoki}},\ }\href
  {https://doi.org/10.1103/PhysRevLett.78.1335} {\bibfield  {journal} {\bibinfo
   {journal} {Phys. Rev. Lett.}\ }\textbf {\bibinfo {volume} {78}},\ \bibinfo
  {pages} {1335} (\bibinfo {year} {1997})}\BibitemShut {NoStop}%
\bibitem [{\citenamefont {Zhu}\ \emph {et~al.}(2011)\citenamefont {Zhu},
  \citenamefont {Levy}, \citenamefont {Ludbrook}, \citenamefont {Veenstra},
  \citenamefont {Rosen}, \citenamefont {Comin}, \citenamefont {Wong},
  \citenamefont {Dosanjh}, \citenamefont {Ubaldini}, \citenamefont {Syers},
  \citenamefont {Butch}, \citenamefont {Paglione}, \citenamefont {Elfimov},\
  and\ \citenamefont {Damascelli}}]{Zhu2011a}%
  \BibitemOpen
  \bibfield  {author} {\bibinfo {author} {\bibfnamefont {Z.-H.}\ \bibnamefont
  {Zhu}}, \bibinfo {author} {\bibfnamefont {G.}~\bibnamefont {Levy}}, \bibinfo
  {author} {\bibfnamefont {B.}~\bibnamefont {Ludbrook}}, \bibinfo {author}
  {\bibfnamefont {C.~N.}\ \bibnamefont {Veenstra}}, \bibinfo {author}
  {\bibfnamefont {J.~A.}\ \bibnamefont {Rosen}}, \bibinfo {author}
  {\bibfnamefont {R.}~\bibnamefont {Comin}}, \bibinfo {author} {\bibfnamefont
  {D.}~\bibnamefont {Wong}}, \bibinfo {author} {\bibfnamefont {P.}~\bibnamefont
  {Dosanjh}}, \bibinfo {author} {\bibfnamefont {A.}~\bibnamefont {Ubaldini}},
  \bibinfo {author} {\bibfnamefont {P.}~\bibnamefont {Syers}}, \bibinfo
  {author} {\bibfnamefont {N.~P.}\ \bibnamefont {Butch}}, \bibinfo {author}
  {\bibfnamefont {J.}~\bibnamefont {Paglione}}, \bibinfo {author}
  {\bibfnamefont {I.~S.}\ \bibnamefont {Elfimov}},\ and\ \bibinfo {author}
  {\bibfnamefont {A.}~\bibnamefont {Damascelli}},\ }\href
  {https://doi.org/10.1103/PhysRevLett.107.186405} {\bibfield  {journal}
  {\bibinfo  {journal} {Phys. Rev. Lett.}\ }\textbf {\bibinfo {volume} {107}},\
  \bibinfo {pages} {186405} (\bibinfo {year} {2011})}\BibitemShut {NoStop}%
\bibitem [{\citenamefont {Benia}\ \emph {et~al.}(2011)\citenamefont {Benia},
  \citenamefont {Lin}, \citenamefont {Kern},\ and\ \citenamefont
  {Ast}}]{Benia2011}%
  \BibitemOpen
  \bibfield  {author} {\bibinfo {author} {\bibfnamefont {H.~M.}\ \bibnamefont
  {Benia}}, \bibinfo {author} {\bibfnamefont {C.}~\bibnamefont {Lin}}, \bibinfo
  {author} {\bibfnamefont {K.}~\bibnamefont {Kern}},\ and\ \bibinfo {author}
  {\bibfnamefont {C.~R.}\ \bibnamefont {Ast}},\ }\href
  {https://doi.org/10.1103/PhysRevLett.107.177602} {\bibfield  {journal}
  {\bibinfo  {journal} {Phys. Rev. Lett.}\ }\textbf {\bibinfo {volume} {107}},\
  \bibinfo {pages} {177602} (\bibinfo {year} {2011})}\BibitemShut {NoStop}%
\bibitem [{\citenamefont {King}\ \emph {et~al.}(2011)\citenamefont {King},
  \citenamefont {Hatch}, \citenamefont {Bianchi}, \citenamefont {Ovsyannikov},
  \citenamefont {Lupulescu}, \citenamefont {Landolt}, \citenamefont {Slomski},
  \citenamefont {Dil}, \citenamefont {Guan}, \citenamefont {Mi}, \citenamefont
  {Rienks}, \citenamefont {Fink}, \citenamefont {Lindblad}, \citenamefont
  {Svensson}, \citenamefont {Bao}, \citenamefont {Balakrishnan}, \citenamefont
  {Iversen}, \citenamefont {Osterwalder}, \citenamefont {Eberhardt},
  \citenamefont {Baumberger},\ and\ \citenamefont {Hofmann}}]{King2011}%
  \BibitemOpen
  \bibfield  {author} {\bibinfo {author} {\bibfnamefont {P.~D.~C.}\
  \bibnamefont {King}}, \bibinfo {author} {\bibfnamefont {R.~C.}\ \bibnamefont
  {Hatch}}, \bibinfo {author} {\bibfnamefont {M.}~\bibnamefont {Bianchi}},
  \bibinfo {author} {\bibfnamefont {R.}~\bibnamefont {Ovsyannikov}}, \bibinfo
  {author} {\bibfnamefont {C.}~\bibnamefont {Lupulescu}}, \bibinfo {author}
  {\bibfnamefont {G.}~\bibnamefont {Landolt}}, \bibinfo {author} {\bibfnamefont
  {B.}~\bibnamefont {Slomski}}, \bibinfo {author} {\bibfnamefont {J.~H.}\
  \bibnamefont {Dil}}, \bibinfo {author} {\bibfnamefont {D.}~\bibnamefont
  {Guan}}, \bibinfo {author} {\bibfnamefont {J.~L.}\ \bibnamefont {Mi}},
  \bibinfo {author} {\bibfnamefont {E.~D.~L.}\ \bibnamefont {Rienks}}, \bibinfo
  {author} {\bibfnamefont {J.}~\bibnamefont {Fink}}, \bibinfo {author}
  {\bibfnamefont {A.}~\bibnamefont {Lindblad}}, \bibinfo {author}
  {\bibfnamefont {S.}~\bibnamefont {Svensson}}, \bibinfo {author}
  {\bibfnamefont {S.}~\bibnamefont {Bao}}, \bibinfo {author} {\bibfnamefont
  {G.}~\bibnamefont {Balakrishnan}}, \bibinfo {author} {\bibfnamefont {B.~B.}\
  \bibnamefont {Iversen}}, \bibinfo {author} {\bibfnamefont {J.}~\bibnamefont
  {Osterwalder}}, \bibinfo {author} {\bibfnamefont {W.}~\bibnamefont
  {Eberhardt}}, \bibinfo {author} {\bibfnamefont {F.}~\bibnamefont
  {Baumberger}},\ and\ \bibinfo {author} {\bibfnamefont {P.}~\bibnamefont
  {Hofmann}},\ }\href {https://doi.org/10.1103/PhysRevLett.107.096802}
  {\bibfield  {journal} {\bibinfo  {journal} {Phys. Rev. Lett.}\ }\textbf
  {\bibinfo {volume} {107}},\ \bibinfo {pages} {1} (\bibinfo {year}
  {2011})}\BibitemShut {NoStop}%
\bibitem [{\citenamefont {Yaji}\ \emph {et~al.}(2016)\citenamefont {Yaji},
  \citenamefont {Harasawa}, \citenamefont {Kuroda}, \citenamefont {Toyohisa},
  \citenamefont {Nakayama}, \citenamefont {Ishida}, \citenamefont {Fukushima},
  \citenamefont {Watanabe}, \citenamefont {Chen}, \citenamefont {Komori},\ and\
  \citenamefont {Shin}}]{Yaji2016}%
  \BibitemOpen
  \bibfield  {author} {\bibinfo {author} {\bibfnamefont {K.}~\bibnamefont
  {Yaji}}, \bibinfo {author} {\bibfnamefont {A.}~\bibnamefont {Harasawa}},
  \bibinfo {author} {\bibfnamefont {K.}~\bibnamefont {Kuroda}}, \bibinfo
  {author} {\bibfnamefont {S.}~\bibnamefont {Toyohisa}}, \bibinfo {author}
  {\bibfnamefont {M.}~\bibnamefont {Nakayama}}, \bibinfo {author}
  {\bibfnamefont {Y.}~\bibnamefont {Ishida}}, \bibinfo {author} {\bibfnamefont
  {A.}~\bibnamefont {Fukushima}}, \bibinfo {author} {\bibfnamefont
  {S.}~\bibnamefont {Watanabe}}, \bibinfo {author} {\bibfnamefont
  {C.}~\bibnamefont {Chen}}, \bibinfo {author} {\bibfnamefont {F.}~\bibnamefont
  {Komori}},\ and\ \bibinfo {author} {\bibfnamefont {S.}~\bibnamefont {Shin}},\
  }\href {https://doi.org/10.1063/1.4948738} {\bibfield  {journal} {\bibinfo
  {journal} {Rev. Sci. Instrum.}\ }\textbf {\bibinfo {volume} {87}},\ \bibinfo
  {pages} {053111} (\bibinfo {year} {2016})}\BibitemShut {NoStop}%
\bibitem [{\citenamefont {Yaji}\ \emph {et~al.}(2018)\citenamefont {Yaji},
  \citenamefont {Harasawa}, \citenamefont {Kuroda}, \citenamefont {Li},
  \citenamefont {Yan}, \citenamefont {Komori},\ and\ \citenamefont
  {Shin}}]{Yaji2018}%
  \BibitemOpen
  \bibfield  {author} {\bibinfo {author} {\bibfnamefont {K.}~\bibnamefont
  {Yaji}}, \bibinfo {author} {\bibfnamefont {A.}~\bibnamefont {Harasawa}},
  \bibinfo {author} {\bibfnamefont {K.}~\bibnamefont {Kuroda}}, \bibinfo
  {author} {\bibfnamefont {R.}~\bibnamefont {Li}}, \bibinfo {author}
  {\bibfnamefont {B.}~\bibnamefont {Yan}}, \bibinfo {author} {\bibfnamefont
  {F.}~\bibnamefont {Komori}},\ and\ \bibinfo {author} {\bibfnamefont
  {S.}~\bibnamefont {Shin}},\ }\href
  {https://doi.org/10.1103/PhysRevB.98.041404} {\bibfield  {journal} {\bibinfo
  {journal} {Phys. Rev. B}\ }\textbf {\bibinfo {volume} {98}},\ \bibinfo
  {pages} {041404} (\bibinfo {year} {2018})}\BibitemShut {NoStop}%
\bibitem [{\citenamefont {Miller}\ \emph {et~al.}(1988)\citenamefont {Miller},
  \citenamefont {Samsavar}, \citenamefont {Franklin},\ and\ \citenamefont
  {Chiang}}]{Miller1988}%
  \BibitemOpen
  \bibfield  {author} {\bibinfo {author} {\bibfnamefont {T.}~\bibnamefont
  {Miller}}, \bibinfo {author} {\bibfnamefont {A.}~\bibnamefont {Samsavar}},
  \bibinfo {author} {\bibfnamefont {G.~E.}\ \bibnamefont {Franklin}},\ and\
  \bibinfo {author} {\bibfnamefont {T.~C.}\ \bibnamefont {Chiang}},\ }\href
  {https://doi.org/10.1103/PhysRevLett.61.1404} {\bibfield  {journal} {\bibinfo
   {journal} {Phys. Rev. Lett.}\ }\textbf {\bibinfo {volume} {61}},\ \bibinfo
  {pages} {1404} (\bibinfo {year} {1988})}\BibitemShut {NoStop}%
\bibitem [{\citenamefont {McMahon}\ \emph {et~al.}(1993)\citenamefont
  {McMahon}, \citenamefont {Miller},\ and\ \citenamefont
  {Chiang}}]{McMahon1993}%
  \BibitemOpen
  \bibfield  {author} {\bibinfo {author} {\bibfnamefont {W.~E.}\ \bibnamefont
  {McMahon}}, \bibinfo {author} {\bibfnamefont {T.}~\bibnamefont {Miller}},\
  and\ \bibinfo {author} {\bibfnamefont {T.~C.}\ \bibnamefont {Chiang}},\
  }\href {https://doi.org/10.1103/PhysRevLett.71.907} {\bibfield  {journal}
  {\bibinfo  {journal} {Phys. Rev. Lett.}\ }\textbf {\bibinfo {volume} {71}},\
  \bibinfo {pages} {907} (\bibinfo {year} {1993})}\BibitemShut {NoStop}%
\bibitem [{\citenamefont {Chiang}(2000)}]{Chiang2000}%
  \BibitemOpen
  \bibfield  {author} {\bibinfo {author} {\bibfnamefont {T.-C.}\ \bibnamefont
  {Chiang}},\ }\href {https://doi.org/10.1016/S0167-5729(00)00006-6} {\bibfield
   {journal} {\bibinfo  {journal} {Surf. Sci. Rep.}\ }\textbf {\bibinfo
  {volume} {39}},\ \bibinfo {pages} {181} (\bibinfo {year} {2000})}\BibitemShut
  {NoStop}%
\bibitem [{\citenamefont {Cercellier}\ \emph {et~al.}(2004)\citenamefont
  {Cercellier}, \citenamefont {Fagot-Revurat}, \citenamefont {Kierren},
  \citenamefont {Reinert}, \citenamefont {Popovi{\'{c}}},\ and\ \citenamefont
  {Malterre}}]{Cercellier2004}%
  \BibitemOpen
  \bibfield  {author} {\bibinfo {author} {\bibfnamefont {H.}~\bibnamefont
  {Cercellier}}, \bibinfo {author} {\bibfnamefont {Y.}~\bibnamefont
  {Fagot-Revurat}}, \bibinfo {author} {\bibfnamefont {B.}~\bibnamefont
  {Kierren}}, \bibinfo {author} {\bibfnamefont {F.}~\bibnamefont {Reinert}},
  \bibinfo {author} {\bibfnamefont {D.}~\bibnamefont {Popovi{\'{c}}}},\ and\
  \bibinfo {author} {\bibfnamefont {D.}~\bibnamefont {Malterre}},\ }\href
  {https://doi.org/10.1103/PhysRevB.70.193412} {\bibfield  {journal} {\bibinfo
  {journal} {Phys. Rev. B}\ }\textbf {\bibinfo {volume} {70}},\ \bibinfo
  {pages} {193412} (\bibinfo {year} {2004})}\BibitemShut {NoStop}%
\bibitem [{\citenamefont {Popovi{\'{c}}}\ \emph {et~al.}(2005)\citenamefont
  {Popovi{\'{c}}}, \citenamefont {Reinert}, \citenamefont {H{\"{u}}fner},
  \citenamefont {Grigoryan}, \citenamefont {Springborg}, \citenamefont
  {Cercellier}, \citenamefont {Fagot-Revurat}, \citenamefont {Kierren},\ and\
  \citenamefont {Malterre}}]{Popovic2005}%
  \BibitemOpen
  \bibfield  {author} {\bibinfo {author} {\bibfnamefont {D.}~\bibnamefont
  {Popovi{\'{c}}}}, \bibinfo {author} {\bibfnamefont {F.}~\bibnamefont
  {Reinert}}, \bibinfo {author} {\bibfnamefont {S.}~\bibnamefont
  {H{\"{u}}fner}}, \bibinfo {author} {\bibfnamefont {V.~G.}\ \bibnamefont
  {Grigoryan}}, \bibinfo {author} {\bibfnamefont {M.}~\bibnamefont
  {Springborg}}, \bibinfo {author} {\bibfnamefont {H.}~\bibnamefont
  {Cercellier}}, \bibinfo {author} {\bibfnamefont {Y.}~\bibnamefont
  {Fagot-Revurat}}, \bibinfo {author} {\bibfnamefont {B.}~\bibnamefont
  {Kierren}},\ and\ \bibinfo {author} {\bibfnamefont {D.}~\bibnamefont
  {Malterre}},\ }\href {https://doi.org/10.1103/PhysRevB.72.045419} {\bibfield
  {journal} {\bibinfo  {journal} {Phys. Rev. B}\ }\textbf {\bibinfo {volume}
  {72}},\ \bibinfo {pages} {045419} (\bibinfo {year} {2005})}\BibitemShut
  {NoStop}%
\bibitem [{\citenamefont {Cercellier}\ \emph {et~al.}(2006)\citenamefont
  {Cercellier}, \citenamefont {Didiot}, \citenamefont {Fagot-Revurat},
  \citenamefont {Kierren}, \citenamefont {Moreau}, \citenamefont {Malterre},\
  and\ \citenamefont {Reinert}}]{Cercellier2006}%
  \BibitemOpen
  \bibfield  {author} {\bibinfo {author} {\bibfnamefont {H.}~\bibnamefont
  {Cercellier}}, \bibinfo {author} {\bibfnamefont {C.}~\bibnamefont {Didiot}},
  \bibinfo {author} {\bibfnamefont {Y.}~\bibnamefont {Fagot-Revurat}}, \bibinfo
  {author} {\bibfnamefont {B.}~\bibnamefont {Kierren}}, \bibinfo {author}
  {\bibfnamefont {L.}~\bibnamefont {Moreau}}, \bibinfo {author} {\bibfnamefont
  {D.}~\bibnamefont {Malterre}},\ and\ \bibinfo {author} {\bibfnamefont
  {F.}~\bibnamefont {Reinert}},\ }\href
  {https://doi.org/10.1103/PhysRevB.73.195413} {\bibfield  {journal} {\bibinfo
  {journal} {Phys. Rev. B}\ }\textbf {\bibinfo {volume} {73}},\ \bibinfo
  {pages} {195413} (\bibinfo {year} {2006})}\BibitemShut {NoStop}%
\bibitem [{\citenamefont {Luh}\ \emph {et~al.}(2008)\citenamefont {Luh},
  \citenamefont {Cheng}, \citenamefont {Tsuei},\ and\ \citenamefont
  {Tang}}]{Luh2008}%
  \BibitemOpen
  \bibfield  {author} {\bibinfo {author} {\bibfnamefont {D.-A.}\ \bibnamefont
  {Luh}}, \bibinfo {author} {\bibfnamefont {C.-M.}\ \bibnamefont {Cheng}},
  \bibinfo {author} {\bibfnamefont {K.-D.}\ \bibnamefont {Tsuei}},\ and\
  \bibinfo {author} {\bibfnamefont {J.-M.}\ \bibnamefont {Tang}},\ }\href
  {https://doi.org/10.1103/PhysRevB.78.233406} {\bibfield  {journal} {\bibinfo
  {journal} {Phys. Rev. B}\ }\textbf {\bibinfo {volume} {78}},\ \bibinfo
  {pages} {233406} (\bibinfo {year} {2008})}\BibitemShut {NoStop}%
\bibitem [{\citenamefont {Forster}\ \emph {et~al.}(2011)\citenamefont
  {Forster}, \citenamefont {Gergert}, \citenamefont {Nuber}, \citenamefont
  {Bentmann}, \citenamefont {Huang}, \citenamefont {Gong}, \citenamefont
  {Zhang},\ and\ \citenamefont {Reinert}}]{Forster2011}%
  \BibitemOpen
  \bibfield  {author} {\bibinfo {author} {\bibfnamefont {F.}~\bibnamefont
  {Forster}}, \bibinfo {author} {\bibfnamefont {E.}~\bibnamefont {Gergert}},
  \bibinfo {author} {\bibfnamefont {A.}~\bibnamefont {Nuber}}, \bibinfo
  {author} {\bibfnamefont {H.}~\bibnamefont {Bentmann}}, \bibinfo {author}
  {\bibfnamefont {L.}~\bibnamefont {Huang}}, \bibinfo {author} {\bibfnamefont
  {X.~G.}\ \bibnamefont {Gong}}, \bibinfo {author} {\bibfnamefont
  {Z.}~\bibnamefont {Zhang}},\ and\ \bibinfo {author} {\bibfnamefont
  {F.}~\bibnamefont {Reinert}},\ }\href
  {https://doi.org/10.1103/PhysRevB.84.075412} {\bibfield  {journal} {\bibinfo
  {journal} {Phys. Rev. B}\ }\textbf {\bibinfo {volume} {84}},\ \bibinfo
  {pages} {075412} (\bibinfo {year} {2011})}\BibitemShut {NoStop}%
\bibitem [{\citenamefont {Kuroda}\ \emph {et~al.}(2018)\citenamefont {Kuroda},
  \citenamefont {Yaji}, \citenamefont {Harasawa}, \citenamefont {Noguchi},
  \citenamefont {Kondo}, \citenamefont {Komori},\ and\ \citenamefont
  {Shin}}]{Kuroda2018a}%
  \BibitemOpen
  \bibfield  {author} {\bibinfo {author} {\bibfnamefont {K.}~\bibnamefont
  {Kuroda}}, \bibinfo {author} {\bibfnamefont {K.}~\bibnamefont {Yaji}},
  \bibinfo {author} {\bibfnamefont {A.}~\bibnamefont {Harasawa}}, \bibinfo
  {author} {\bibfnamefont {R.}~\bibnamefont {Noguchi}}, \bibinfo {author}
  {\bibfnamefont {T.}~\bibnamefont {Kondo}}, \bibinfo {author} {\bibfnamefont
  {F.}~\bibnamefont {Komori}},\ and\ \bibinfo {author} {\bibfnamefont
  {S.}~\bibnamefont {Shin}},\ }\href {https://doi.org/10.3791/57090} {\bibfield
   {journal} {\bibinfo  {journal} {J. Vis. Exp.}\ }\textbf {\bibinfo {volume}
  {136}},\ \bibinfo {pages} {e57090} (\bibinfo {year} {2018})}\BibitemShut
  {NoStop}%
\bibitem [{sup()}]{supple}%
  \BibitemOpen
  \href@noop {} {\bibinfo  {journal} {See Supplemental Material at [URL] for
  the details of experimental methods and DFT calculations and SARPES results
  of the surface state}\
  }\BibitemShut {NoStop}%
\end{thebibliography}

%

\end{document}